# Continuous-variable Quantum Key Distribution protocols with eight-state discrete modulation

A. Becir[1, 2], F. A. A. El-Orany[2,3], M. R. B. Wahiddin [1, 2]

[1] *Faculty of Science, International Islamic University of Malaysia (IIUM),*

*P.O. Box 141, 25710 Kuantan,*

*Pahang Darul Makmur, Malaysia*

[2] *Information Security Cluster, MIMOS Berhad, Technology Park Malaysia,*

*57000 Kuala Lumpur, Malaysia*

*3 Department of Mathematics and Computer Science, Faculty of*

*Science, Suez Canal University, Ismailia, Egypt*

**Abstract:** We propose a continuous variable quantum key distribution protocol based on discrete modulation of eight-state coherent states. We present a rigorous security proof against the collective attacks considering both of realistic lossy, noisy quantum channels, imperfect detector efficiency, and detector electronic noise. This protocol shows high tolerance against excess noise and it is promising to achieve hundreds distance long in optical fiber base.

Electronic address: ahmed.becir@mimos.my

## I.Introduction

Quantum properties of the physical systems may realize provably secure communication between two legitimate parties. The main task of the so-called quantum key distribution (QKD) protocols is to achieve secret-key sharing using imperfect semi classical signals and devices [1]. However, since the first demonstration of quantum teleportation using continuous variable (CV) [2], the QKD protocols based on quantum CV systems via coherent states have also been proposed recently for achieving the secret key sharing [3–7]. The main idea in CV QKD is that the legitimate receiver (Bob) of the transmitted signal measures one of the conjugate quadratures randomly; but an exception is found [7].

The security of the Gaussian CV-QKD protocol with the homodyne detection has been proven against individual Gaussian eavesdropping attacks, using either direct [3] or reverse [4] reconciliation. Moreover, the security proof of this protocol against general individual or finite-size attacks [7], and general collective attacks [9-11] has been verified. Also, the security of the Gaussian CV-QKD protocol with the heterodyne detection has been discussed in [12-15]. Precisely, the bounds of Eve's accessible information in the case of individual Gaussian attacks have been developed in [ 12], and later improved in [13, 14], while the case of collective attacks has been analyzed in [15]. Recently, the unconditional security of both the homodyne and the heterodyne protocols has been proven [16].

In spite of the CV QKD shows many advantages over the discrete variable QKD such as high key rate, it is still restricted to a small distance. So far it is experimentally demonstrated over 30Km [11] and theoretically possible over 50Km [17]. The reason of this comes from excess noise produced by the quantum channel and the difficulty of applying the error correction code [11, 17].

Working with high signal-to-noise ratio (SNR) needs a very good reconciliation, otherwise the secret key rate goes to zero [11, 17]. Unfortunately, even with the best codes available today (low-density parity check LDPC codes [18] or turbo codes [19]), still there is no hopes to extend the distance over 50 Km. On the other hand, working with low SNR may increases the range of the protocol [17]. Nonetheless, maintaining good reconciliation efficiency at very low SNR is even more difficult. There are some interesting algebraic properties of $\mathbb{R}^8$ can be useful which may enhance the practical distance up to 50 km [17].

Recently, a new protocols using discrete modulation have been developed [20] and experimentally implemented [21]. These protocols are theoretically promising to achieve hundreds of Km and shows high tolerance against the excess noise generated by the quantum channel. Their property is that they always generate less than bit for each pulse and for any practical distance. This property makes them more suitable to work over the long distance. In fact many bits for each pulse can be achieved using the Gaussian modulation over short distance. After several kilometers the Gaussian modulation will generate less than bit for each pulse as well and the two modulations become comparable. The discrete modulation is more robust against the excess noise and can achieve longer distance while the Gaussian modulation key rate drops below zero after tens of kilometers. It is worth mentioning that an interesting new reconciliation algorithm has been developed for direct and reverse reconciliation [20] as well. This algorithm is promising to achieve more than 80% efficiency even with low SNR [20].

In this paper, we develop new protocol based on the eight coherent states (eight-state protocol). In this protocol Alice sends one of the eight coherent states $|\alpha e^{ik\pi/4}\rangle$ with $k \in \{0,1,2,\ldots,7\}$ to Bob with equal probability. Bob will measure the received states either via homodyne or heterodyne detections. For this protocol we calculate the secret key rate of the collective attacks. We consider the realistic case where the losses and excess noise of the quantum channels can be controlled by the eavesdropper Eve. We take into account also the imperfection of detection, and electronic noise (or thermal noise) generated by the homodyne (heterodyne) detection circuit.

This paper is prepared in the following ordered. In section II, we introduce the protocol of CV-QKD using 8-state coherent discrete modulation. In sections III we review some notation and assumptions. In section IV, we derive corresponding expressions for the secret information rate in the presence of Eve collective attacks. The performance of the protocol in a realistic practical setup in fibre optics base is given in section V. The main results are summarized in section VI.

## II.Eight-State Protocol

In this section we describe the eight-state protocol and give some mathematical justification which will be used in the rest of the paper. More illustratively, we develop the form of the state $|\Psi_8\rangle$, which can be used in an entanglement-based version of the protocol. Also we deduce the covariance matrix for this state.

The eight-state protocol can be described as follows. Alice sends one of the eight displaced coherent states $|\beta_1\rangle = |\alpha\rangle$, $|\beta_2\rangle = |\alpha e^{i\pi/4}\rangle$, $|\beta_1^*\rangle = |\alpha e^{i\pi/2}\rangle$, $|\beta_2^*\rangle = |\alpha e^{i3\pi/4}\rangle$, $|-\beta_1\rangle = |\alpha e^{i\pi}\rangle$, $|-\beta_2\rangle = |\alpha e^{i5\pi/4}\rangle$, $|-\beta_1^*\rangle = |\alpha e^{i3\pi/2}\rangle$, and $|-\beta_2^*\rangle = |\alpha e^{i7\pi/4}\rangle$ with equal probability (, i.e. 1/8) to Bob. Information about the distribution of these eight states in the phase space is shown in figure 1. Bob chooses to measure one of the two quadratures randomly (homodyne case) or both quadratures simultaneously (heterodyne case). Once the quantum transmission phase of the communication has ended, Alice and Bob proceed with classical data processing procedures,

which include a reconciliation algorithm to extract an identical chain of bits from their correlated continuous data, and a standard privacy amplification process to derive a final secret key from this chain. The reconciliation is direct when Alice's data is used as a reference for establishing the key and reverse when the reference is Bob's data. Reverse reconciliation has been shown to offer a great advantage in QKD system performance [9], therefore calculations in this paper have been performed for this case. Direct reconciliation expressions can be derived using similar tools as the ones presented here.

Now we give the mathematical treatment for $|\Psi_8\rangle$. In the first step, Bob receives a mixed state represented by the density matrix $\rho_8$, which has the form:

$$\rho_8 = \frac{1}{8}(|\beta_1\rangle\langle\beta_1 + |\beta_2\rangle\langle\beta_2| + |\beta_1^*\rangle\langle\beta_1^*| + |\beta_2^*\rangle\langle\beta_2^*| + |-\beta_1\rangle\langle-\beta_1| + |-\beta_2\rangle\langle-\beta_2|$$

$$+|-\beta_1^*\rangle\langle-\beta_1^*| + |-\beta_2^*\rangle\langle-\beta_2^*|), \tag{1}$$

$$= \lambda_0|\phi_0\rangle\langle\phi_0| + \lambda_1|\phi_1\rangle\langle\phi_1| + \cdots + \lambda_7|\phi_7\rangle\langle\phi_7|, \tag{2}$$

where

$$\begin{cases} \lambda_{0(4)} = \frac{1}{4}e^{-\alpha^2}\left[\cosh(\alpha^2) + \cos(\alpha^2) \pm 2\cos\left(\frac{\alpha^2}{\sqrt{2}}\right)\cosh\left(\frac{\alpha^2}{\sqrt{2}}\right)\right], \\ \lambda_{1(5)} = \frac{1}{4}e^{-\alpha^2}\left[\sinh(\alpha^2) + \sin(\alpha^2) \pm \sqrt{2}\cos\left(\frac{\alpha^2}{\sqrt{2}}\right)\sinh\left(\frac{\alpha^2}{\sqrt{2}}\right) \pm\sqrt{2}\sin\left(\frac{\alpha^2}{\sqrt{2}}\right)\cosh\left(\frac{\alpha^2}{\sqrt{2}}\right)\right], \\ \lambda_{2(6)} = \frac{1}{4}e^{-\alpha^2}\left[\cosh(\alpha^2) - \cos(\alpha^2) \pm 2\sin\left(\frac{\alpha^2}{\sqrt{2}}\right)\sinh\left(\frac{\alpha^2}{\sqrt{2}}\right)\right], \\ \lambda_{3(7)} = \frac{1}{4}e^{-\alpha^2}\left[\sinh(\alpha^2) - \sin(\alpha^2) \mp \sqrt{2}\cos\left(\frac{\alpha^2}{\sqrt{2}}\right)\sinh\left(\frac{\alpha^2}{\sqrt{2}}\right) \pm\sqrt{2}\sin\left(\frac{\alpha^2}{\sqrt{2}}\right)\cosh\left(\frac{\alpha^2}{\sqrt{2}}\right)\right] \end{cases}, \tag{3}$$

and

$$|\phi_k\rangle = \frac{e^{-\frac{\alpha^2}{2}}}{\sqrt{\lambda_k}}\sum_{n=0}^{\infty}\frac{\alpha^{(8n+k)}}{\sqrt{(8n+k)!}}|8n+k\rangle, \quad \text{for } k \in \{0,1,2,\dots,7\}. \tag{4}$$

Applying the annihilation operator $\hat{a}$ to $|\phi_k\rangle$ gives:

$$\hat{a}\,|\phi_k\rangle = \alpha\frac{\sqrt{\lambda_{k-1}}}{\sqrt{\lambda_k}}\,|\phi_{k-1}\rangle, \quad \text{for } k \in \{1,2,\dots,7\}, \text{ and} \tag{5}$$

$$\hat{a}\,|\phi_0\rangle = -\alpha\frac{\sqrt{\lambda_7}}{\sqrt{\lambda_0}}\,|\phi_7\rangle. \tag{6}$$

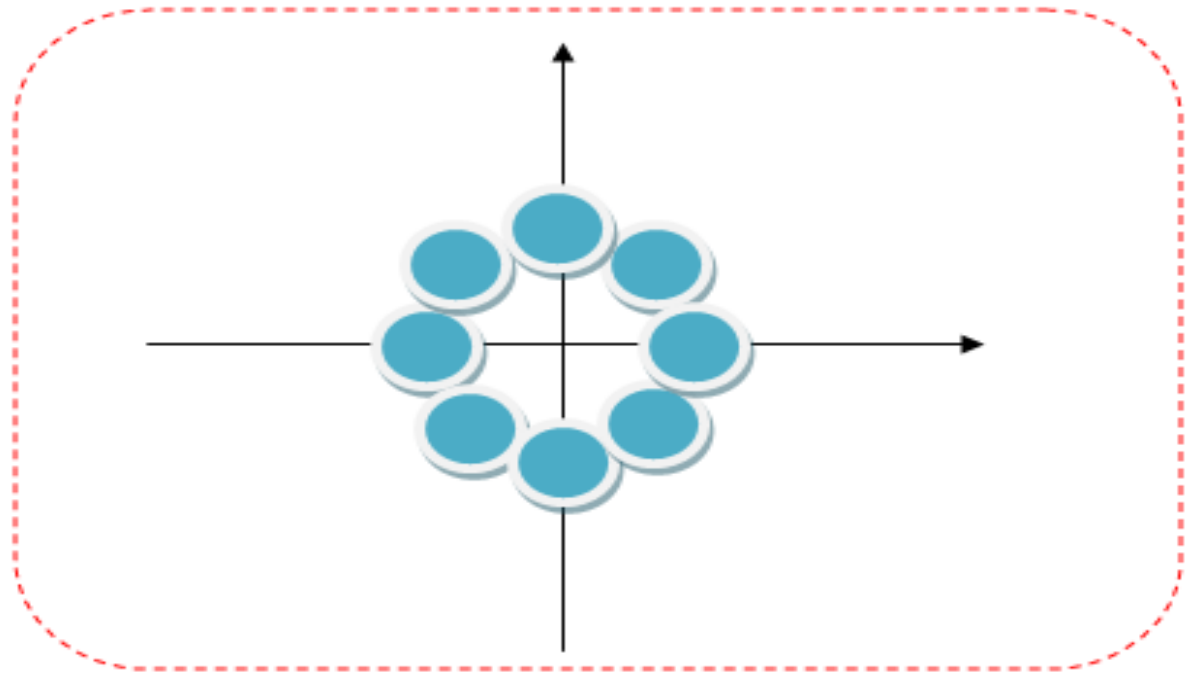

Figure 1: (Color online.) Encoding schemes used for the eight-state protocol

Now we consider for the state $\rho_8$ the following purification

$$|\Psi_8\rangle = \sum_{k=0}^{7} \sqrt{\lambda_k}\, |\phi_k\rangle|\phi_k\rangle. \tag{7}$$

This state can be rewritten as

$$|\Psi_8\rangle = 1/4(|\psi_0\rangle|\beta_1\rangle + |\psi_1\rangle|-\beta_2^*\rangle + |\psi_2\rangle|-\beta_1^*\rangle + |\psi_3\rangle|-\beta_2\rangle + |\psi_4\rangle|-\beta_1\rangle$$
$$+|\psi_5\rangle|\beta_2^*\rangle + |\psi_6\rangle|\beta_1^*\rangle + |\psi_7\rangle|\beta_2\rangle), \tag{8}$$

where

$$|\psi_k\rangle = \frac{1}{2}\sum_{m=0}^{7} e^{\frac{i(4k+1)m\pi}{4}}\, |\phi_m\rangle, \qquad \text{for } k \in \{0,1,2,\dots,7\}. \tag{9}$$

It is evident that states (9) are orthogonal non-Gaussian states. The bipartite state (8) is a good example for the entanglement based scheme, in which Alice performs projective measurement on one of the set $\{|\psi_0\rangle\langle\psi_0|, |\psi_1\rangle\langle\psi_1|, |\psi_2\rangle\langle\psi_2|, \dots |\psi_7\rangle\langle\psi_7|\}$ to the first half of $|\Psi_8\rangle$ and that she therefore projects the second half on one of the eight coherent states $|\beta_1\rangle$, $|\beta_2\rangle,\dots,|-\beta_1^*\rangle$ with equal probabilities .

Now we are in the position to evaluate the covariance matrix $\Gamma_8$ of the bipartite state $|\Psi_8\rangle$ . Using symmetry arguments $\Gamma_8$ has the following form:

$$\Gamma_8 = \begin{pmatrix} X\mathbb{1}_2 & Z_8\sigma \\ Z_8\sigma & Y\mathbb{1}_2 \end{pmatrix}, \tag{10}$$

with

$$\begin{cases} X = \langle \Psi_8 | 2a^+a + 1 | \Psi_8 \rangle, \\ Y = \langle \Psi_8 | 2b^+b + 1 | \Psi_8 \rangle\,, \\ Z_8 = \langle \Psi_8 | ab + a^+b^+ | \Psi_8 \rangle, \end{cases} \tag{11}$$

where $a$, $a^+$and $b$, $b^+$are the annihilation and creation operators related to Alice and Bob modes, respectively. Therefore, the covariance matrix elements $X$, $Y$and $Z_8$ then read:

$$X = Y = 1 + 2\alpha^2 = 1 + V_A, \text{ and} \tag{12}$$

$$Z_8 = 2\alpha^2 \sum_{k=0}^{7} \frac{\lambda_{k-1}{}^{3/2}}{\sqrt{\lambda_k}} = V_A \sum_{k=0}^{7} \frac{\lambda_{k-1}{}^{3/2}}{\sqrt{\lambda_k}} \tag{13}$$

We conclude this section by comparing the behavior of $Z_8$with that of $Z_G$ $of$ the Gaussian modulation protocol ($Z_G = \sqrt{V^2 - 1}\, with\, V = V_A + 1$) and $Z_4$ of the four-state discrete modulation protocol [20]. Information about this is shown in figure 2. It is clear that for variance $V_A \leq 1$, $Z_8$ and $Z_G$ are almost indistinguishable, meaning that in this regime, one has the Holevo bound $S_8(B, E) \approx S_G(B, E)$. Therefore, the security bounds for the Gaussian modulation can be carried here as well.

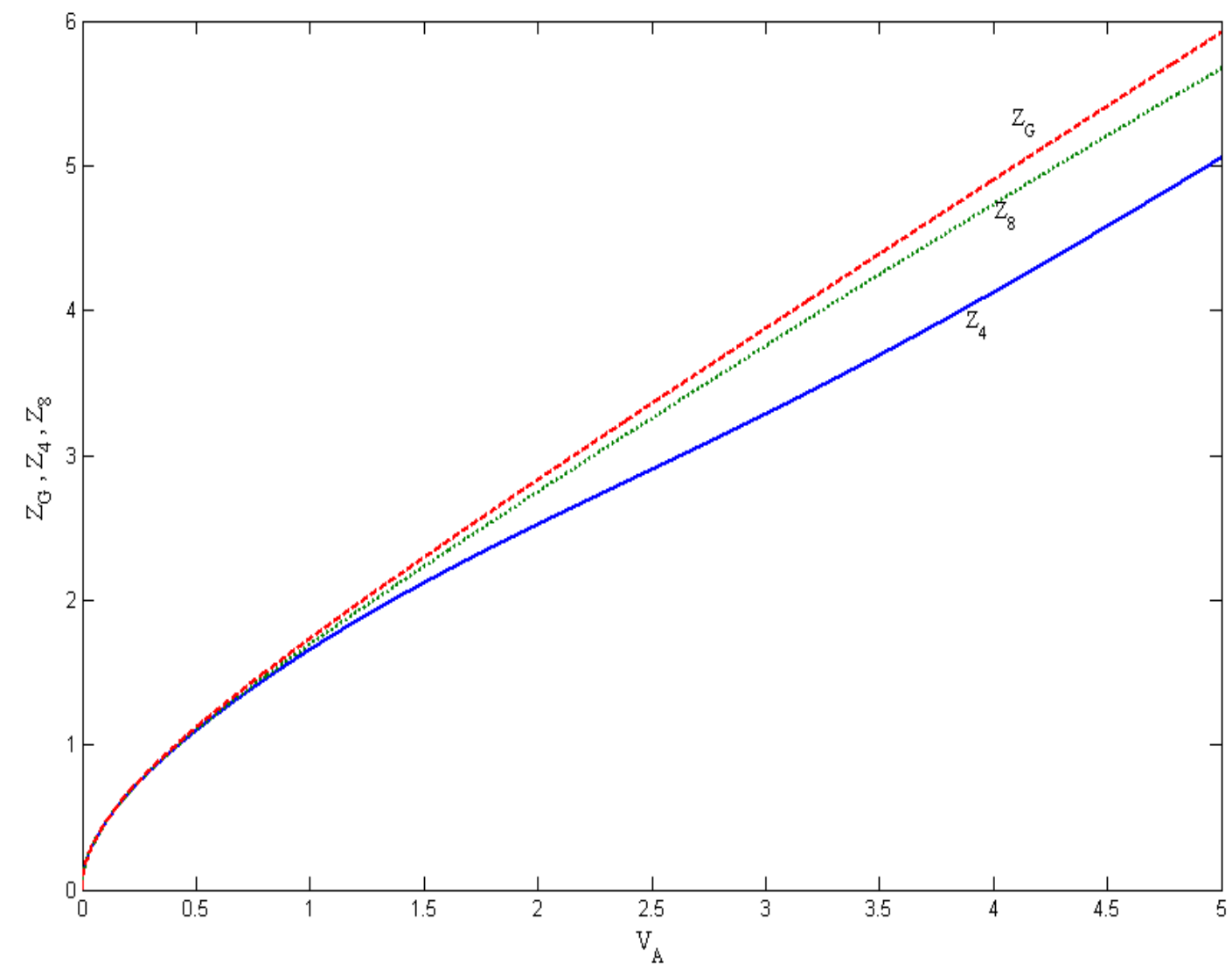

Figure2: Comparison between the Correlations: $Z_G = \sqrt{V^2 - 1}$ (Gaussian modulation), $Z_4$ (4-state discrete modulation) and $Z_8$ (8-state discrete modulation).

## III.Notations and assumptions

In this section we describe the parameters of the eight-state protocol. These parameters will be used to study the collective attacks. In the execution of the protocol Alice and Bob use quantum channels and detectors (homodyne or heterodyne). We assume that the channel features transmission efficiency $T$ and excess noise $\varepsilon$. These parameters result in a noise variance at Bob's station as $(1 + T\,\varepsilon)N_0$, where $N_0$ refers to shot-noise limit. The total channel-added noise referred to the channel input, expressed in shot noise units, is defined as $\chi_{line} = 1/T - 1 + \varepsilon$. Furthermore, we assume the detector in the Bob's station is characterized by the efficiency $\eta$ (due to the losses) and a noise $\varepsilon_{ele}$ (due to the thermal noise introduced by the electrical circuit). The total added noise referred to the homodyne (heterodyne), expressed in shot-noise units, is defined as $\chi_{hom} = [(1-\eta) + \varepsilon_{ele}]/\eta$ for homodyne and $\chi_{het} = [1 + (1-\eta) + 2\varepsilon_{ele}]/\eta$ for heterodyne detection. Therefore, the total noise referred to the channel input can then be expressed $\chi_T = \chi_{line} + \chi_{\mathrm{hom}\,(het)}/T$.

We conclude this section by mentioning that the eight-state protocol is a prepare-and-measure scheme. This is based on the fact that Alice prepares and sends one of the eight displaced coherent states and Bob measures these station in his work station. In this respect, this protocol is equivalent to the entanglement-based scheme, which is shown in figure 3. Bob's detector inefficiency is modelled by a beam-splitter with transmission $\eta$, while its electronic noise $\varepsilon_{ele}$ is modelled by an EPR state of variance $N$, one half of which is entering the second input port of the beam-splitter, as shown in figure 3. The variance $N$ is chosen to obtain an appropriate expression of $\chi_{\mathrm{hom}\,(het)}$. For homodyne detection, $N = \eta\, \chi_{hom}/(1-\eta)$, and for heterodyne $N = (\eta\, \chi_{het} - 1)/(1-\eta)$.

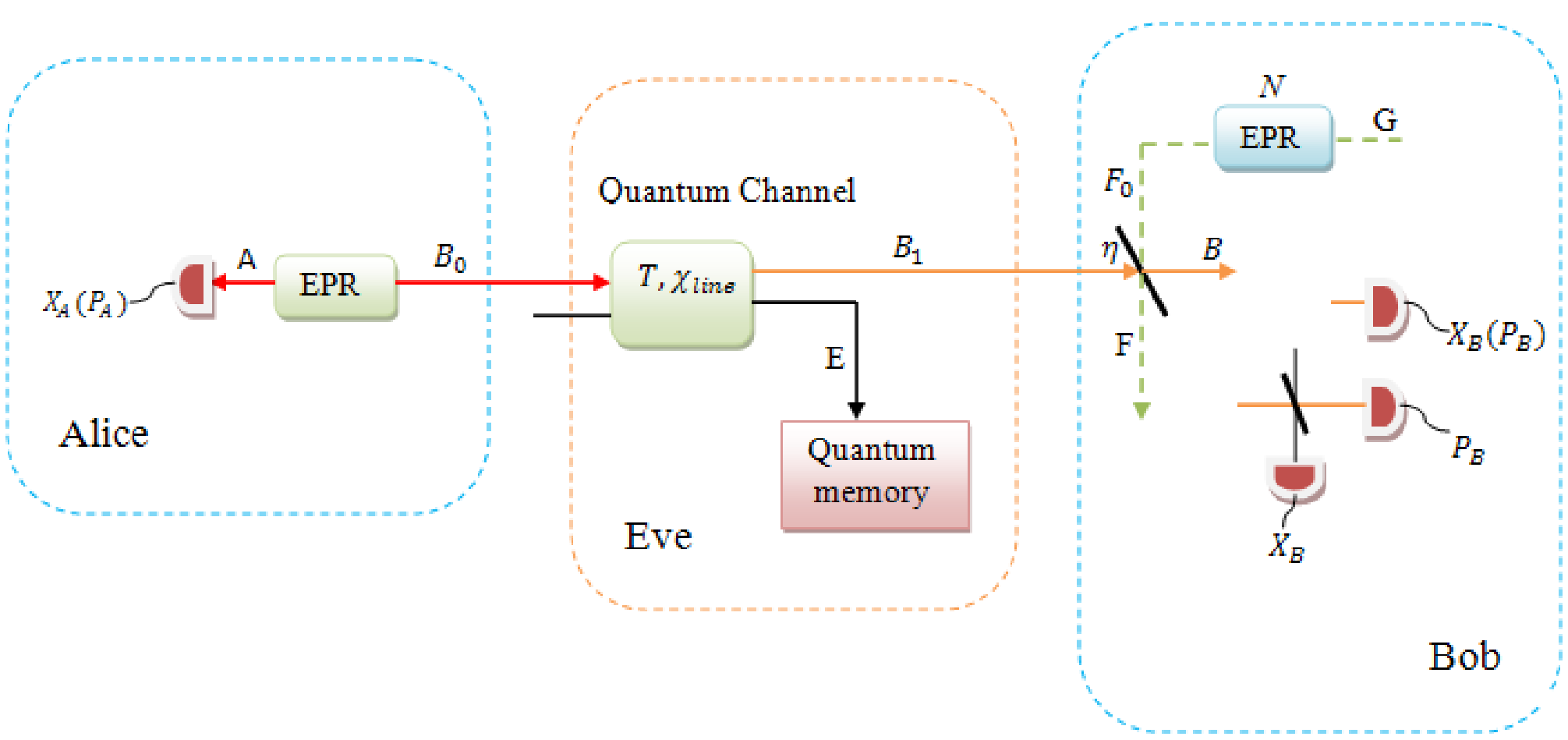

Figure 3: Entanglement-based scheme of 8-state discrete modulation CVQKD protocol with homodyne or heterodyne detection. The transmission $T$ and channel-added noise $\chi_{line}$ are controlled by Eve, who does not have access to Bob's detection apparatus.

## IV.Collective attacks:

In this section, we use the results in the preceding sections to discuss the collective attacks for the protocol under consideration. In the collective attacks, Eve interacts individually with each pulse. We assume that it is allowed for her to wait till the entire classical process to be ended before performing the collective measurement on her stored ancillae. In this attack, the maximum information accessible to Eve is limited by the Holevo bound $\chi_{BE}$ [22]. In the reverse reconciliation and under the realistic case, the Holevo secret key rate is given by:

$$\Delta I^{Holevo} \geq \beta I_{AB} - \chi_{BE}, \tag{14}$$

where $\beta$ is the reconciliation algorithm efficiency and $I_{AB}$ is the mutual information between Alice and Bob that is given by [11, 23]:

$$I_{AB} = \frac{1}{2}\log\frac{V_B}{V_{B/A}} = \frac{1}{2}\log\frac{V+\chi_T}{1+\chi_T}, \tag{15}$$

where $V_{B/A}$ is the conditional variance of Alice based on Bob measurement $V_{B/A} = \langle {X_B}^2\rangle - \langle X_A X_B\rangle^2/\langle {X_A}^2\rangle$ . Note that in the case of heterodyne detection the mutual information $I_{AB}$ is double of the one in the expression (15). The Holevo bound $\chi_{BE}$ is given by [9-11]:

$$\chi_{BE} = S\left(\rho_{AB_1}\right) - S\left(\rho_{AFG}^{x_E}\right) = \sum_{i=1}^{2} G\left(\frac{\lambda_i - 1}{2}\right) - \sum_{j=3}^{5} G\left(\frac{\lambda_j - 1}{2}\right), \tag{16}$$

where $G(x) = (x+1)\log(x+1) - x\log x$ , $\lambda_{1,2}$ are the symplectic eiegenvalues of the covariance matrix $\gamma_{AB_1}$, and $\lambda_{3,4,5}$ are the symplectic eiegenvalues of the covariance matrix $\gamma_{AFG}^{x_E}$ after Bob`s projective measurement (see Fig. 3).

The covariance matrix $\gamma_{AB_1}$ depends on the system including Alice and the channel, and it's given by:

$$\gamma_{AB_1} = \begin{bmatrix} \gamma_A & \sigma_{AB_1} \\ \sigma_{AB_1} & \gamma_{B_1} \end{bmatrix} = \begin{bmatrix} V\mathbb{I}_2 & \sqrt{T}Z_8\sigma_z \\ \sqrt{T}Z_8\sigma_z & T(V+\chi_{line})\mathbb{I}_2 \end{bmatrix}, \tag{17}$$

The symplectic eigenvalues $\lambda_{1,2}$ are then given by:

$$\lambda_{1,2} = \sqrt{\frac{1}{2}\left(\Delta \pm \sqrt{\Delta^2 - 4\mathrm{D}}\right)}, \text{with} \tag{18}$$

$$\begin{cases} \Delta = \det\, \gamma_{AB_1} \\ \mathrm{D} = \det\, \gamma_A + \det\, \gamma_{B_1} + 2\det\, \sigma_{AB_1} \end{cases} \tag{19}$$

The entropy $S\left(\rho_{AFG}^{x_E}\right)$ is determined from the symplectic eigenvalues $\lambda_{3,4,5}$ of the covariance matrix $\gamma_{AFG}^{x_E}$ after Bob's projective measurement. The matrix $\gamma_{AFG}^{x_E}$ is written as:

$$\gamma_{AFG}^{x_E} = \gamma_{AFG} - {\sigma_{AFGB}}^T H\, \sigma_{AFGB}, \tag{20}$$

where $H$ stands for the symplectic matrix which represents the homodyne (heterodyne) measurement on mode $B$. In the former case $H_{hom} = (X^T \gamma_B X)^{MP}$, where $X = \begin{pmatrix} 1 & 0 \\ 0 & 0 \end{pmatrix}$ and MP stands for Moore-Penrose Pseudo-inverse of a matrix for homodyne and $H_{het} = (\gamma_B + \mathbb{I}_2)^{-1}$ for heterodyne detection. The matrices $\gamma_B$, $\gamma_{AFG}$ and $\sigma_{AFGB}$ can be derived from the decomposition of the covariance matrix:

$$\gamma_{AFGB} = \begin{bmatrix} \gamma_{AFG} & {\sigma_{AFGB}}^T \\ \sigma_{AFGB} & \gamma_B \end{bmatrix} \tag{21}$$

The above matrix can be derived with appropriate rearrangement of rows and columns of the matrix describing the system $\gamma_{ABFG}$, which is given by:

$$\gamma_{AFGB} = (Y^{BS})^T \left(\gamma_{AB_1} \oplus \gamma_{F_0 G}\right) Y^{BS} \tag{22}$$

The matrix $Y^{BS}$describes the beamsplitter transformation that models the inefficiency of the detector, which acts on modes $B_1$and $F_0$ and it has the form:

$$Y^{BS} = \mathbb{I}_2 \oplus Y_{B_1 F_0}^{BS} \oplus \mathbb{I}_2 \tag{23}$$

Now we can proceed to calculate the symplectic eigenvalues $\lambda_{3,4,5}$ of the three-mode matrix (20). The symplectic eigenvalues $\lambda_{3,4,5}$ are given by

$$\lambda_{3,4} = \sqrt{\frac{1}{2}\left(\mathrm{A} \pm \sqrt{\mathrm{A}^2 - 4\mathrm{B}}\right)}, \qquad \lambda_5 = 1 \tag{24}$$

where for homodyne case [11] we have

$$A_{hom} = \frac{\Delta\chi_{hom} + V\sqrt{D} + T(V + \chi_{line})}{T(V + \chi_T)}, \tag{25}$$

$$B_{hom} = \sqrt{D}\,\frac{\mathrm{V} + \sqrt{D}\chi_{hom}}{T(V + \chi_T)}. \tag{26}$$

For heterodyne case [14] we have

$$A_{het} = \frac{\Delta\chi_{hom}{}^2 + D + 1 + 2\chi_{hom}\left[V\sqrt{D} + T(V + \chi_{line}) + 2TZ_8\right]}{T^2(V + \chi_T)^2}, \tag{27}$$

$$B_{het} = \left(\frac{\mathrm{V} + \sqrt{D}\chi_{het}}{T(V + \chi_T)}\right)^2. \tag{28}$$

The parameters $\Delta$ and $D$ are given by equations (19). Based on the equations (14), (18-19) and (24-28), we can calculate the Holevo bound $\chi_{BE}$ and thus derive the Holevo secret information rate $\Delta I^{Holevo} = \beta I_{AB} - \chi_{BE}$. In the following section we investigate the performance of our eight-state protocol based on the discrete modulation in the realistic setup.

## V.Application to practical systems:

In this section, we apply the result derived in sections II and IV to practical QKD system. Precisely, we calculate the secret key generation rate as a function of distance for fibre-optic implementations of CV-QKD protocol in the existence of collective eavesdropping attacks. This will be done for the homodyne and heterodyne detections.

The numerical simulations of the secret information rate are shown in the figures $4(a, b)$ and $5(a, b)$ for the four-state and eight-state protocols, respectively.

In the numerical treatment, the channel transmission efficiency is given as $T = 10^{-\mu L/10}$, where $\mu = 0.2 db/Km$ is the loss coefficient for the standard optical fibres, and $L$ is the length of the fibre optics. In most practical CVQKD protocols, the efficiency of detection has the value $\eta = 0.6$ (60%), the electrical noise introduced by the homodyne detection circuits is $\varepsilon_{ele} = 0.05$ and the excess noise introduced by the channel (standard fibre optics) is $\varepsilon = 0.005$ (in shot-noise units) [11]. Here we assume that Eve intercepts the quantum channel and produces extra excess noise. Thus, we study the tolerance of our QKD protocol against different values of excess noise, in particular, we take $\varepsilon = 0.005$ (in absence of Eve), $\varepsilon = 0.01$ and $\varepsilon = 0.02$ (in shot-noise unit). The reconciliation of algorithm is fixed to $\beta = 0.8$, which is the actual values for the homodyne detection protocols [20, 21].

It is worth pointing out that we have considered the same reconciliation efficiency for both configurations, i.e. homodyne and heterodyne detections. Therefore, both of them seem to generate same key rate. Nonetheless, in practice the heterodyne detection can give better reconciliation than that of the heterodyne detection.

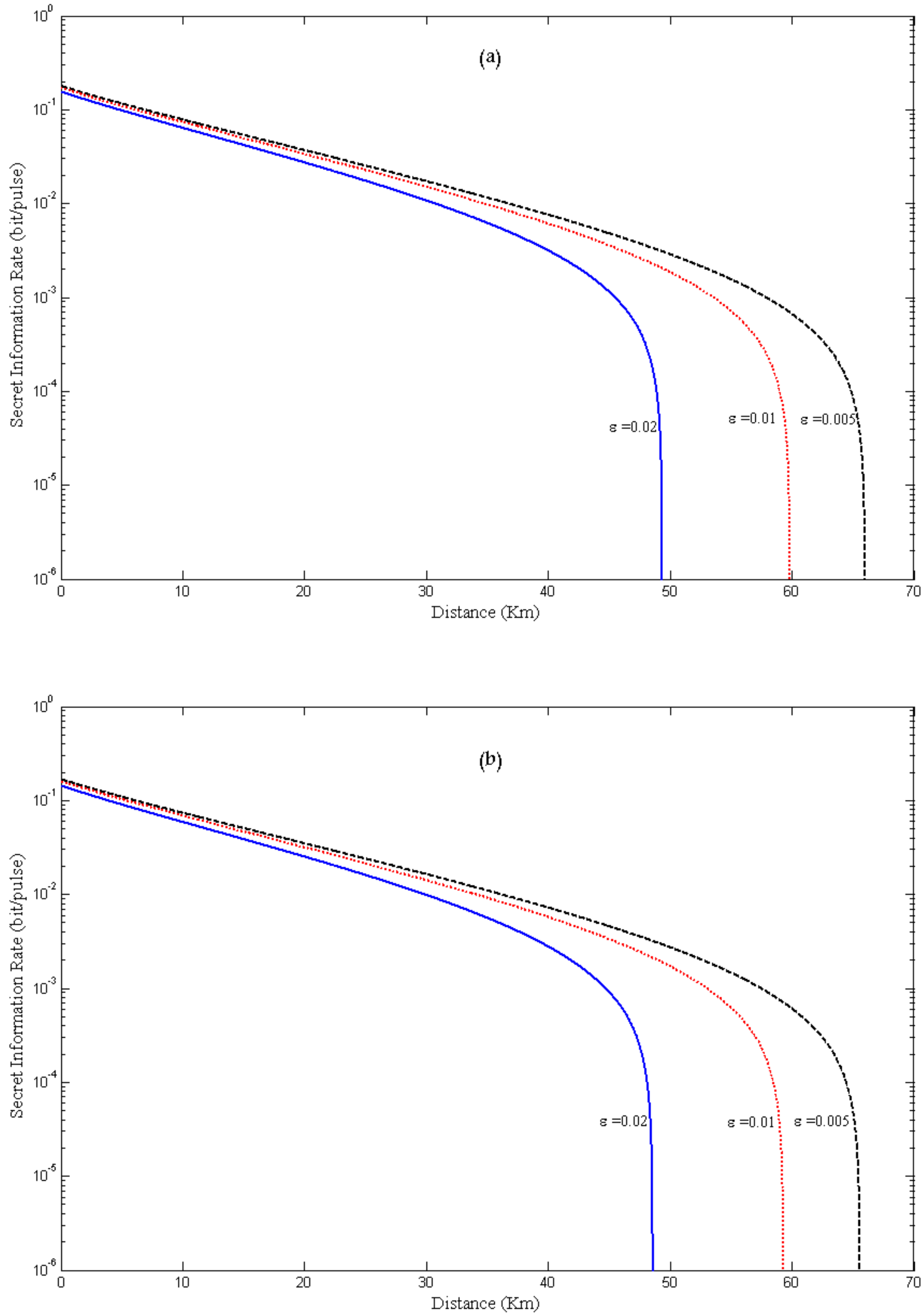

Figure 4: Secret key rate of the 4-state protocol for realistic reconciliation efficiency of 80% and a quantum efficiency of Bob's detection equal to 0.6 with thermal noise $\varepsilon_{ele} = 0.05$ (in shot-noise unit).$V_A$ = 1. (a) for homodyne and (b) for heterodyne.

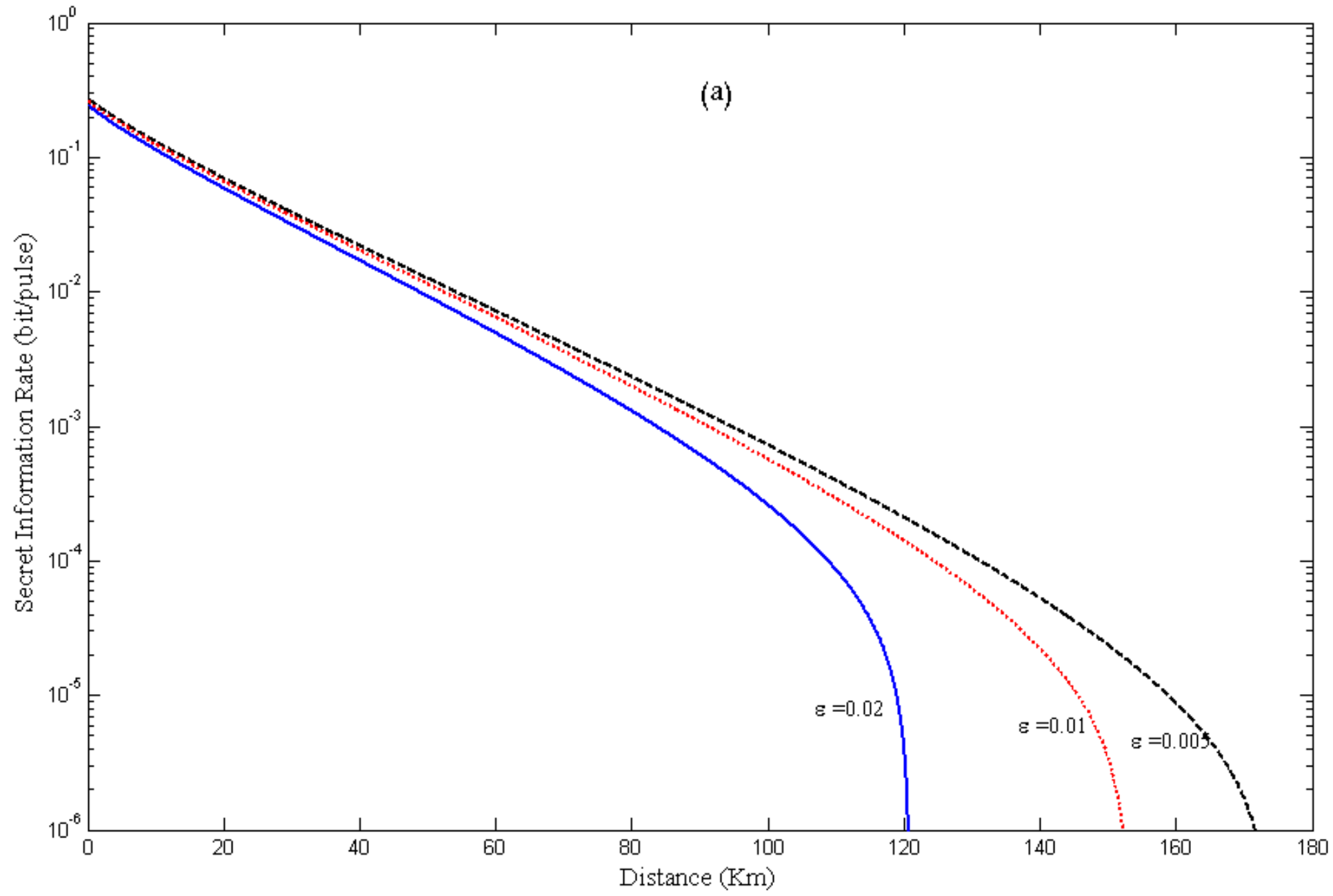

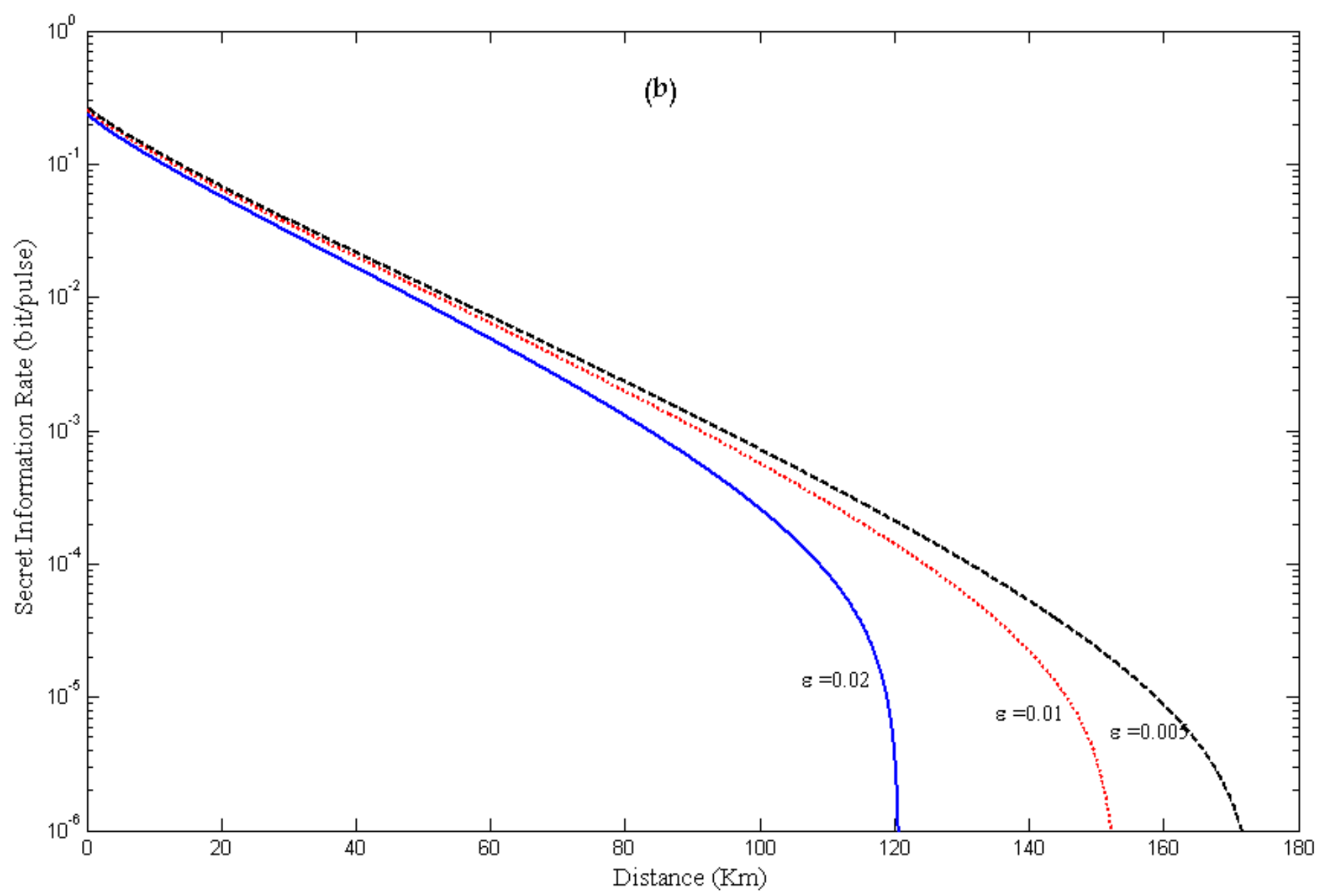

Figure 5: Secret key rate of the eight-state protocol for realistic reconciliation efficiency of 80% and a quantum efficiency of Bob's detection equal to 0.6 with thermal noise $\varepsilon_{ele} = 0.05$ (in shot-noise unit). $V_A = 1$. (a) for homodyne and (b) for heterodyne.

The comparison between Figs. 4 and 5 leads to that the eight-state protocol performance is much better than the four-state protocol. More illustratively, it distributes the secret keys over longer distances, offers higher key rate and tolerates much excess noise. The tolerance of excess noise is extremely important since the excess noise is not always fixed to $\varepsilon = 0.005$ (in shot-noise unit), but it can vary from one experiment to other one. From figures 5, it is obvious that the eight-state protocol can distribute positive key over 100 Km even in the presence of high excess noise. Thus, this protocol is extremely promising to gain hundreds of Kilometers in fiber optics base with current equipments as well as the reconciliation algorithm introduced in [20].

**Conclusion**

In this paper, we have developed the eight-state protocol. We have studied the security of the protocol against the collective attacks taken into account the realistic lossy, noisy quantum channels, imperfect detector efficiency, and detector electronic noise. The protocol shows high tolerance against excess noise and can achieve hundreds of kilometers distance long in optical fiber base with the current technology.